\documentclass[utf8]{FrontiersinVancouver} 
\usepackage{url,hyperref,lineno,microtype}
\usepackage{caption}
\usepackage[onehalfspacing]{setspace}
\usepackage{multirow}
\usepackage{comment}
\usepackage{braket}
\def\keyFont{\fontsize{8}{11}\helveticabold }

\def\firstAuthorLast{Hebborn {et~al.}} 

\def \Address{$^{1}$ Université Paris-Saclay, CNRS/IN2P3, IJCLab, 91405 Orsay, France \\
$^{2}$Facility for Rare Isotope Beams, Michigan State University, East Lansing, Michigan 48824, USA \\
$^{3}$Department of Physics and Astronomy, Michigan State University, East Lansing, Michigan 48824, USA  }
\def\corrAuthor{C. Hebborn}
\def\corrEmail{hebborn@ijlab.in2p3.fr}

\begin{document}
\newcommand {\nc} {\newcommand}
\nc {\IR} [1]{\textcolor{red}{#1}}
\nc {\IB} [1]{\textcolor{blue}{#1}}
\nc{\lsim}{\raisebox{-0.13cm}{~\shortstack{$<$ \\[-0.07cm] $\sim$}}~}
\onecolumn
\firstpage{1}

\title[Systematic study of the propagation of  uncertainties to transfer observables]{Systematic study of the propagation of  uncertainties to transfer observables} 
{\huge \bf \noindent Systematic study of the propagation of  uncertainties to transfer observables}

\noindent{\LARGE \bf C. Hebborn\,$^{1,2,3*}$ and F. M. Nunes\,$^{2,3}$}

\noindent{\Large {$^{1}$ Université Paris-Saclay, CNRS/IN2P3, IJCLab, 91405 Orsay, France \\
	$^{2}$Facility for Rare Isotope Beams, Michigan State University, East Lansing, Michigan 48824, USA \\
	$^{3}$Department of Physics and Astronomy, Michigan State University, East Lansing, Michigan 48824, USA  }\\

\noindent Correspondence: \\
C. Hebborn \\
hebborn@ijclab.in2p3.fr}\\


\begin{abstract}
	A systematic study of parametric uncertainties in transfer reactions is performed using the recently developed uncertainty quantified global optical potential  (KDUQ).
	We consider reactions on the doubly-magic spherical nucleus $^{48}$Ca  and explore the dependence of the predicted  $(d,p)$  angular distribution uncertainties at different beam energies and for different properties of the final single-particle state populated by the reaction. Our results show that correlations between the uncertainties associated with the bound state potential and with the optical potentials may be important for correctly determining the uncertainty in the transfer cross sections (in our case, these do not add in quadrature). In general, we find small uncertainties in the predicted transfer observables: half-width of the 68\% credible interval is roughly  $5-10$\%, which is comparable to the experimental error on the transfer data. Finally, our results show that the relative magnitude of the parametric uncertainty in transfer observables  increases with the beam energy and does not depend strongly on the properties of the final state.

	\tiny
	\keyFont{\section{Keywords:} nuclear reactions, optical model, single-nucleon transfer reactions, uncertainty quantification}
	
\end{abstract}


\section{Introduction}
\label{intro}

Transfer reactions are widely used in nuclear experimental studies, either for extracting astrophysical information that cannot be obtained directly or for studying    properties of the nucleus of interest (e.g. Refs.~\cite{kay2013,Avila2015,walter2019,salathe2020,
	PhysRevC.103.055809,Hammachereview,KayA15,PhysRevLett.129.122501,jones2022,dpfission,hebbornavila}). However, reaction theory is essential to interpret transfer reactions measurements \cite{annual2020,Reviewdp}. The properties we wish to extract from  transfer reactions, such as spectroscopic factors (SF), asymptotic normalization coefficients (ANC) or neutron capture rates, depend strongly on the normalization of the transfer cross section while the  model used to describe the reaction carries uncertainties that affect the normalization~\cite{annual2020}. Thus, for a reliable interpretation of  transfer measurements, it is crucial that we understand the uncertainties associated with the theory.

In this work we  focus on  $(d,p)$  reactions. The preferred model for interpreting  single-nucleon $(d,p)$ transfer reactions is the adiabatic wave approximation (ADWA) \cite{JT74,NLAT}. This model has the advantage that it includes deuteron breakup non-perturbatively, without increasing the computational cost as compared to the Distorted Wave Born Approximation (DWBA). It has also been shown to fare well compared to the state-of-the-art models in the field \cite{nunes2011, upadhyay2012}. In ADWA, the input interactions are nucleon optical potentials, in addition to  potentials that bind the deuteron and the final state. From all the studies performed so far,  optical potentials are the dominant source of uncertainty in  ADWA predictions for transfer $(d,p)$ cross sections. It is important not only to quantify those uncertainties but also understand how they may change with beam energy and specific properties of the final state being produced.

In the last few years, many studies have been performed to quantify the uncertainty in $(d,p)$ reactions using Bayesian statistics \cite{lovell2018,king2018,lovell2021,catacora2023}. These studies use optical potentials fitted for one projectile-target combination, at a specific beam energy,  constrained with a single or a couple data sets. Typically, proton or neutron elastic scattering data at the relevant beam energies are used in a Bayesian calibration to obtain parameter posterior distributions for the optical potentials. The uncertainties in the $(d,p)$ cross sections for the reaction are then obtained by sampling these posterior distributions, which are then propagated using the ADWA framework. Uncertainties obtained in \cite{lovell2018,king2018,lovell2021,catacora2023} are large, in part due to the choice of the likelihood function \cite{pruitt2024}.  By propagating uncertainties from each optical potential independently, no correlations  between the neutron and proton optical potentials in the entrance channel are included, nor between those and the proton optical potential in the exit channel. Recent work on charge-exchange reactions has shown that the inclusion of such correlations can make a significant difference in the uncertainty estimate~\cite{whitehead2022,smith2024}.

Moreover, Ref.~\cite{catacora2023} studies the uncertainties coming from the single-particle potential that binds the neutron in the final state in  $(d,p)$ reactions, in addition to the uncertainties in the optical potential.
By constraining the geometry of the binding potential with the asymptotic normalization coefficient, one can greatly reduce the uncertainties (see Fig. 2 of Ref.~\cite{catacora2023}). If the ANC squared is poorly known the uncertainties in the $(d,p)$ cross section are very large. If the ANC squared is known to say 10\% then the uncertainties in the cross section are greatly reduced.

The recent development of an uncertainty-quantified global optical model (KDUQ) \cite{KDUQ}, based on the original work \cite{KD03}, provides another avenue to study the uncertainties in  reactions.  Propagating the parametric uncertainties from KDUQ to reaction observables has been done  for specific cases (e.g. for knockout and transfer \cite{prl2023,prl2023err} and charge-exchange \cite{smith2024}).
In general, the uncertainties due to optical potentials in reaction observables can be influenced by many different details of the reaction process~\cite{KOUQ22,op2023}. Due to strong non-linearities in the reaction model, it is important to study these uncertainties more systematically to understand the impact of correlations in optical potential parameters and whether there are general features that emerge. KDUQ provides a unified effective framework to perform this study.

This work is a systematic study of the uncertainties associated with the optical potentials in transfer $(d,p)$ observables. We use $^{48}$Ca$(d,p)^{49}$Ca to set up the problem and consider a range of  beam energies as well as a variety of  final bound states with different properties. In Section~\ref{sec2}, we briefly describe the reaction and statistical models used.  In Section~\ref{sec:data}, we compare the  predictions obtained with KDUQ to existing elastic scattering and transfer data, to establish our framework for a realistic case. In the rest of Section \ref{sec:results}, we vary beam energies and final bound state properties (separation energy, angular momentum, nodes, etc) and analyze the dependencies of the resulting uncertainties. Section~\ref{sec:ccl} presents the conclusions of this work.

\section{A brief summary of the theory used}
\label{sec2}

\subsection{Reaction Theory}
\label{sec:reactiontheory}

The finite-range ADWA \cite{JT74} starts out by considering a full three-body picture for the transfer reaction $A(d,p)B$. As detailed in Ref.~\cite{JT74}, it uses Weinberg states to then simplify the T-matrix to
\begin{equation}\label{T_ad}
T^{ADWA} = \bra{\phi_{nA}\chi_{pB}^{(-)}}V_{np}\ket{\phi_{np}\chi_{d}^{ad}}\;.
\end{equation}
In Eq.~\eqref{T_ad}, $\phi_{np}$ and $\phi_{nA}$ correspond to the deuteron bound state and single-particle state of the final nucleus B, $V_{np}$ is the neutron-proton interaction and $\chi_{pB}$ is the outgoing distorted wave of the proton relative to the final nucleus $B$, obtained with the optical potential $U_{pB}$ at the energy of the outgoing proton.
The adiabatic distorted wave $\chi^{ad}_{d}$ is generated from the effective adiabatic potential:
\begin{equation}\label{V_ad}
U^{\text{eff}}_{Ap}= -\bra{\phi_o}V_{np}(U_{nA}+U_{pA})\ket{\phi_o},
\end{equation}
where $U_{nA}$ and $U_{pA}$ are the nucleon optical potentials between neutron/proton and the target evaluated at half the deuteron incoming energy. The wave function $\phi_o$ is the first Weinberg basis state, which is directly proportional to the deuteron bound state \cite{JT74}. The T-matrix of Eq.~\eqref{T_ad} assumes that the remnant term ($U_{nA}-U_{pB}$) is negligible. In ADWA calculations, the sources of parametric uncertainties are  therefore the optical potentials used to generate the scattering states and the single-particle potentials used to model bound states.

More details about the  ADWA and  how to obtain numerical solutions for bound and  scattering states can be found in Ref.~\cite{ThompsonNunesBook}. In this work, we use  the  code {\sc NLAT}~\cite{NLAT} to perform all ADWA transfer calculations {and the code {\sc FRESCO}~\cite{Fresco} to perform the elastic scattering calculations}. 

\subsection{Statistical model}
\label{sec:stat}

As mentioned in Section~\ref{intro}, we use the global optical  potentials KDUQ \cite{KDUQ} for all nucleon-nucleus interactions needed in the reaction model of Section~\ref{sec:reactiontheory}, which  is valid for  $ 24 \leq A \leq 209$ and 1~keV~$\leq E \leq 200 $ MeV. In this work, we chose the democratic version of KDUQ which weighs every data point equally\footnote{{We also consider the federal version of KDUQ \cite{KDUQ}, in which each data type is given equal weight on the overall likelihood. Using the federal KDUQ, we obtained  transfer angular distributions exhibiting similar uncertainties as  the the ones obtained with the democratic KDUQ.}}.  By performing a Bayesian calibration using  a large set of reaction data (including nucleon elastic scattering angular distributions and analyzing powers, neutron total cross sections and proton reaction cross sections, all on stable nuclei), the authors of KDUQ obtained  parameter posterior distributions and  correlations for the 46 parameters of their global optical model. In this work, we use the 416 samples of their posterior distributions, published in Ref.~\cite{KDUQ}, to compute the uncertainties in the transfer cross sections. We  quantify the uncertainty in the transfer angular distribution in terms of the relative half-width of the $1\sigma$ credible interval at the peak of the angular distributions (corresponding to a scattering angle~$\theta_{max}$):
\begin{eqnarray}
\varepsilon_{68\%}&=&\frac{\sigma^{{68\%}}_{max}(\theta_{max})-\sigma^{{68\%}}_{{avg}}(\theta_{max})} {\sigma_{{avg}}(\theta_{max})}\label{halfwidth_eq}\\
\text{with}\quad  \sigma^{{68\%}}_{{avg}}(\theta_{max})&=&\frac{\sigma^{{68\%}}_{max}(\theta_{max})+\sigma^{{68\%}}_{min}(\theta_{max})}{2} 
\end{eqnarray}

The test case we  focus on  is based on the $^{48}$Ca$(d,p)^{49}$Ca(g.s) reaction.
Our choice is mainly motivated by the availability of elastic and transfer data. Moreover, since no $^{48}$Ca data were included in the KDUQ corpus, the analysis performed in this work share similar challenges as the ones on transfer data populating exotic nuclei. 
All optical potentials are taken \textit{consistently} from KDUQ and evaluated at the relevant energies, i.e., all  three nucleon-nucleus optical potentials are derived  from the same KDUQ sample.  This consistent treatment hence includes correlations between the various optical potential parameters. For describing the final state of the neutron, we    take a standard radius and diffuseness $r_R=1.25$ fm and $a_R=0.65$ fm  (STD) or we take the geometry of the real part of the KDUQ interaction (KDUQ-real). In both cases, we adjust the depth to reproduce the neutron separation energy and the parameters in the spin orbit term for the final bound-state potential are fixed: the depth is $V_{so}=6$~MeV, the radius is $r_{so}=1.25$~fm and the diffuseness is $a_{so}=0.65$~fm.
In the physical $^{49}$Ca ground state,  the neutron is in a $1p3/2$ orbital,  bound by $S_n=5.146$~MeV. We  also consider  in Section~\ref{sec:spproperties} other configurations for the final state being populated in the  $(d,p)$  reaction. 

{It must be underlined that  KDUQ posterior distributions contain no constrain from bound state data. Our assumption is to consider that the geometry of the mean field generated by the target nucleus does not change considerably from bound to scattering states. This is consistent with the KDUQ assumption, since it uses an energy-independent parametrization for the  radius and diffuseness of the real term, and these parameters are well constrained by the Bayesian calibration. When using 416 samples for the real radius and diffuseness of KDUQ and refitting the real depth to reproduce the correct neutron separation energy (KDUQ-real). {The resulting ANC-squared $\mathcal{C}^2$ distribution (not shown) is  slightly multimodal and is well constrained: its half-width is about 5\%. Interestingly, the value predicted by KDUQ-real  $\mathcal{C}^2=28.6\pm1.3$~fm$^{-1}$ is consistent with the value $\mathcal{C}^2=32.1\pm3.2$~fm$^{-1}$~\cite{PhysRevC.77.051601} determined from the analysis of various $^{48}$Ca$(d,p)^{49}$Ca(g.s.) datasets at various energies, i.e.,  2~MeV, 13~MeV, 19~MeV, 30~MeV and 56~MeV~\cite{RAPAPORT1972337,ca-dp-data,UOZUMI1994123}. This surprising agreement   seems fortuitous, as ANCs for states of $^{87}$Kr, $^8$B and $^{10}$B predicted by KDUQ do not match the values extracted from transfer and breakup data~\cite{PhysRevC.99.054625,PhysRevC.106.024607,PhysRevC.61.064616}.}
	The KDUQ-real posteriors obtained in this way will also be used to quantify uncertainties from the neutron singe-particle interaction in $\phi_{nA}$.} %

\section{Results}
\label{sec:results}

\begin{figure}[h!]
	\centering
	\includegraphics[width=0.48\linewidth]{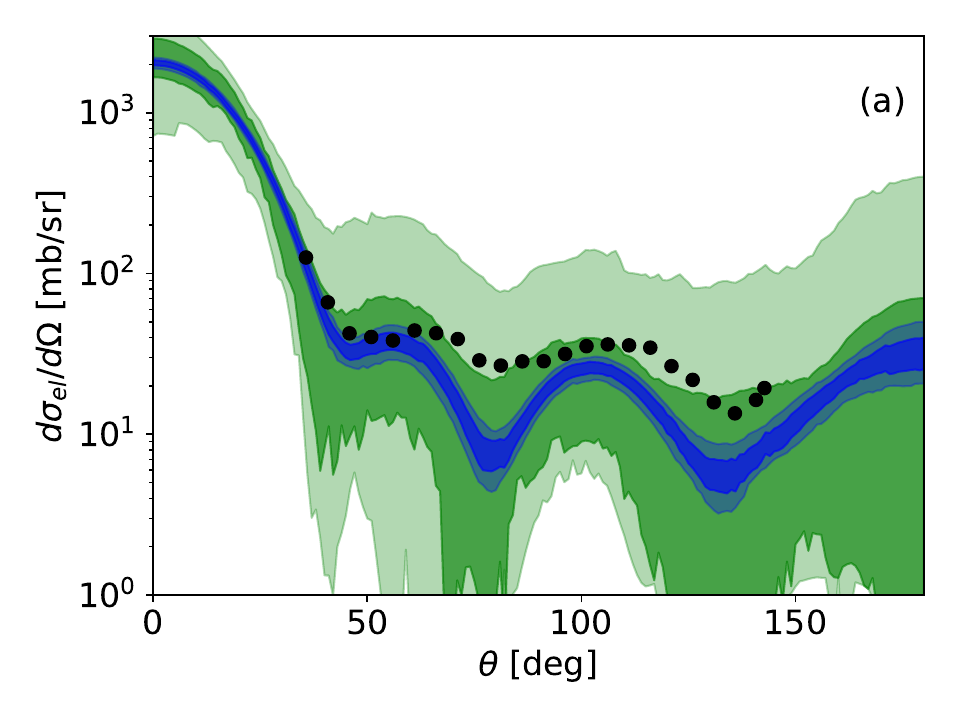}
	\includegraphics[width=0.48\linewidth]{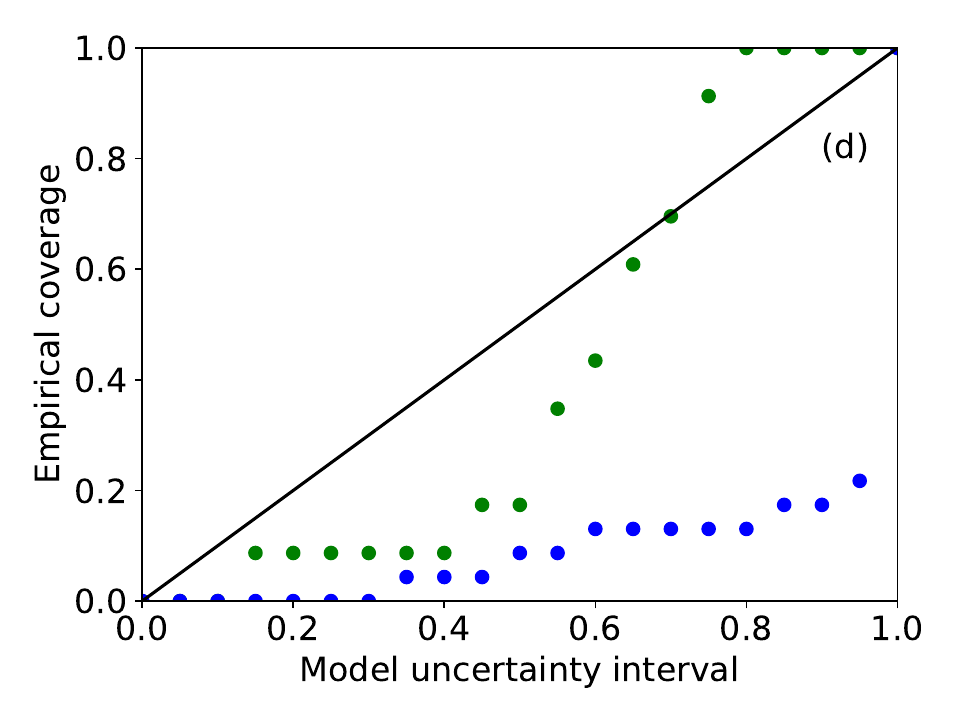}\\
	\includegraphics[width=0.48\linewidth]{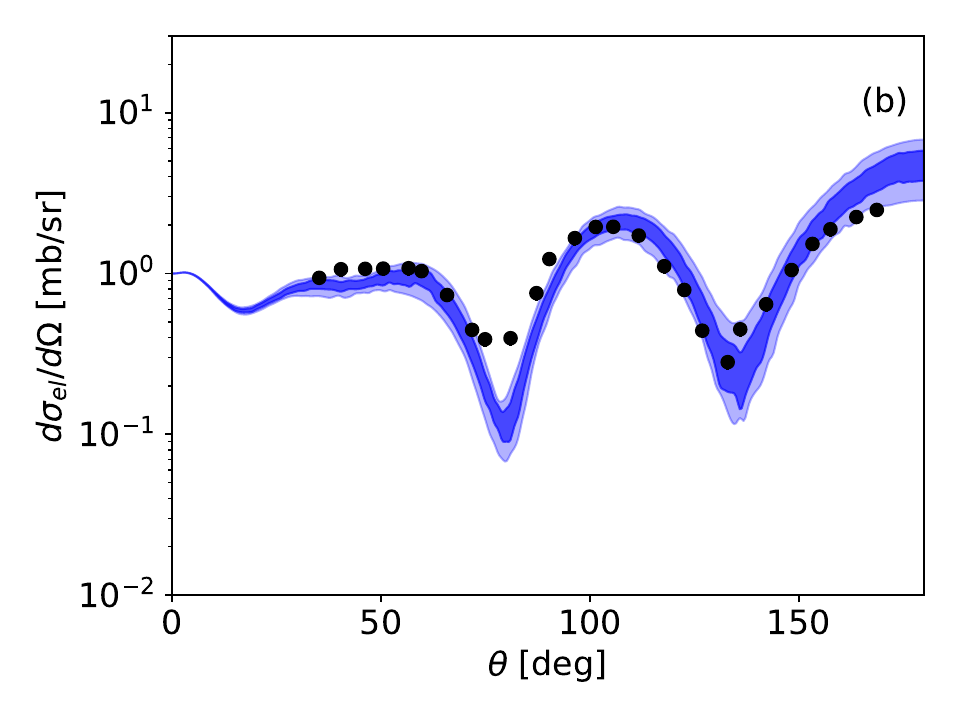}
	\includegraphics[width=0.48\linewidth]{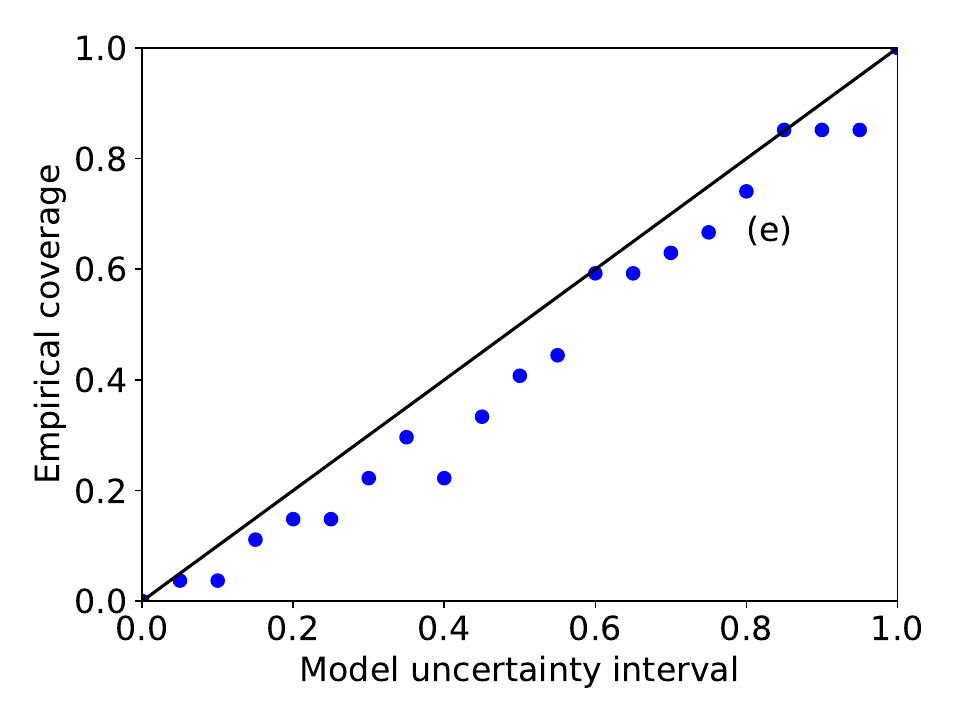}\\
	\includegraphics[width=0.48\linewidth]{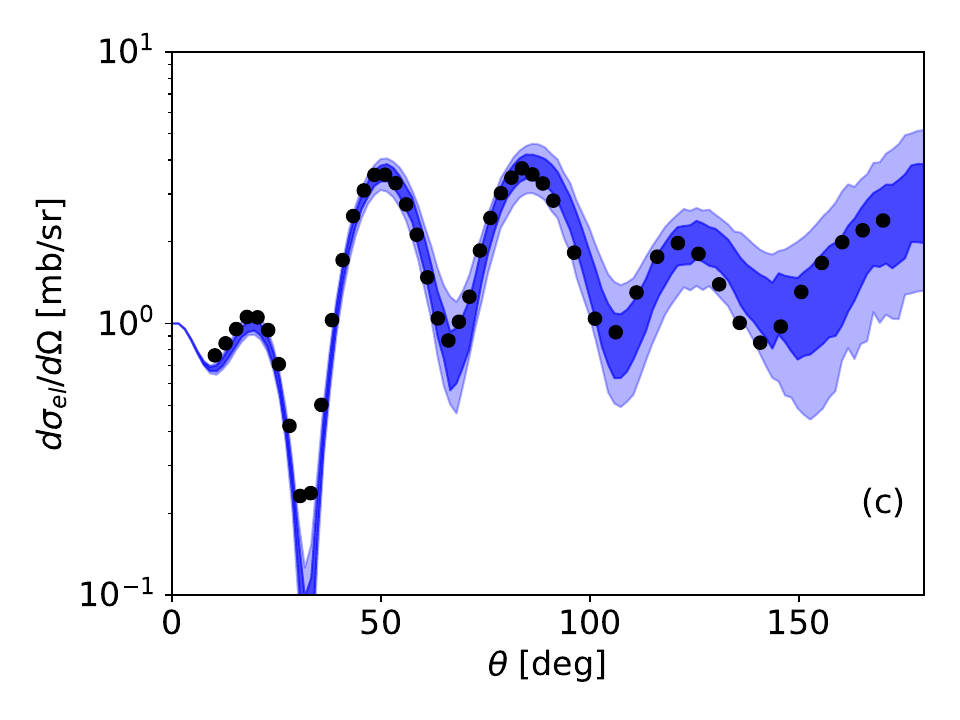}
	\includegraphics[width=0.48\linewidth]{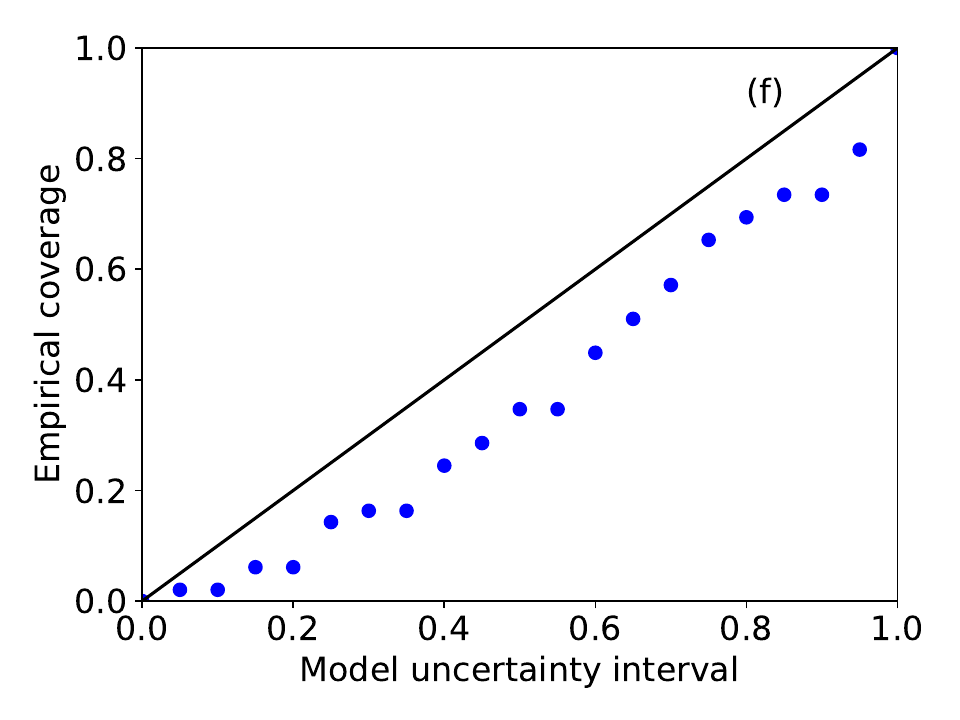}\\
	\caption{Angular distributions for the elastic scattering of (a) $n$+$^{48}$Ca @ 12 MeV, (b) $p$+$^{48}$Ca @ 14 MeV and (c) $p$+$^{48}$Ca @ 25 MeV. The dark and light  shaded blue  bands correspond respectively to the 68\% and  95\% credible intervals  obtained with optical potentials derived from the KDUQ posterior distribution.  The green bands are obtained with rescaled KDUQ posterior distributions (referred as KDUQ-n) in the text. These predictions are compared with data from Refs.\cite{dataeln48Ca,datap48Ca14,datap48Ca25}. (d-f) Corresponding empirical coverage for elastic-scattering data. \\ }
	\label{fig:datael}
\end{figure}
\

\subsection{The physical $^{48}$Ca$(d,p)^{49}$Ca(g.s) reaction}
\label{sec:data}

Optical potentials are determined primarily from  observables that are sensitive to the on-shell T-matrix. Transfer observables are also sensitive to properties of the T-matrix off-shell. It is thus not guaranteed, even if the optical potentials reproduce the corresponding elastic channels, that they  describe the transfer data. Using the reaction $^{48}$Ca$(d,p)^{49}$Ca(g.s)  at 19 MeV, for which there is data \cite{ca-dp-data}\footnote{The uncertainties of this data set are not clearly reported. We consider a 10\% relative error per    data point, which is a typical error for transfer data on stable nuclei.}, we compare predictions using KDUQ with the corresponding data to assess the quality of the uncertainty quantification.
We select nucleon elastic scattering data that is close to the  energies relevant to the transfer reaction of interest ($E_n=9.5$ MeV and $E_p=9.5$ MeV in the entrance channel and $E_p=21.9$ MeV in the exit channel) and compare the credible intervals generated from the KDUQ parameter posteriors with the actual data (with the quoted experimental error bars)~\cite{dataeln48Ca,datap48Ca14,datap48Ca25}. The corresponding angular distributions are shown in Figs.~\ref{fig:datael}: (a) neutron elastic scattering for $E_{lab}=12$ MeV, (b) proton elastic scattering for $E_{lab}=14$ MeV, (c) proton elastic scattering for $E_{lab}=25$ MeV. Note that the KDUQ global optical potential was not fitted on these data sets. The dark (light) shade corresponds to the $68$\% ($95$\%) credible intervals\footnote{{The x\% credible intervals are computed as the smallest   interval that include x\% of the cross section predicted by the 416 samples of KDUQ.}}.

The empirical coverage provides a sanity check for  uncertainty quantification. Our empirical coverages {for a x\% model uncertainty  are calculated as the number of data points, including a x\%   experimental error, that fall into the theoretical x\% }credible interval divided by the total number of data points in an angular distribution. These are shown in Figs.~\ref{fig:datael}(d-f) for the corresponding three elastic scattering examples. In an ideal situation, the predicted empirical coverage should line up with the black diagonal line\footnote{All results in blue in Figs.~\ref{fig:datael} correspond to the uncertainties from KDUQ when the data protocol is  democratic  \cite{KDUQ}.}.
In our calculations for proton elastic scattering, empirical coverages calculated at the high-confidence level are only slightly underestimated. However, for  neutron scattering,  there is a severe mismatch. This suggests that, in this case, the error on the data~\cite{dataeln48Ca} and/or on the KDUQ parameters are seriously under-reported.  To include unaccounted-for uncertainties, we have  inflated the width of the  posterior distributions of the depths, radii and diffuseness of the neutron-target potential,  so that we can reproduce the correct empirical coverage specifically for $1\sigma$. We do this by approximating the parameter distributions of the neutron-target potential $U_{nA}$  to a multivariate Gaussian. We then rescale the covariance matrix by a factor: it turns out that we need to rescale these by $38$, effectively rescaling uncertainties by a factor of $\sqrt{38}\sim 6$. Note that such approach, although simplistic, allows to keep the correlations between the optical potentials parameters informed by the large KDUQ corpus. We  refer to this as KDUQ-n. Replacing KDUQ by KDUQ-n for $U_{nA}$ results in the green bands in Fig.~\ref{fig:datael}(a) and the green dots in Fig.~\ref{fig:datael}(d).  As can be seen, the empirical coverage obtained for the $68$\% is now exactly $68$\%.

\begin{figure}[t!]
	\centering
	\includegraphics[width=0.48\linewidth]{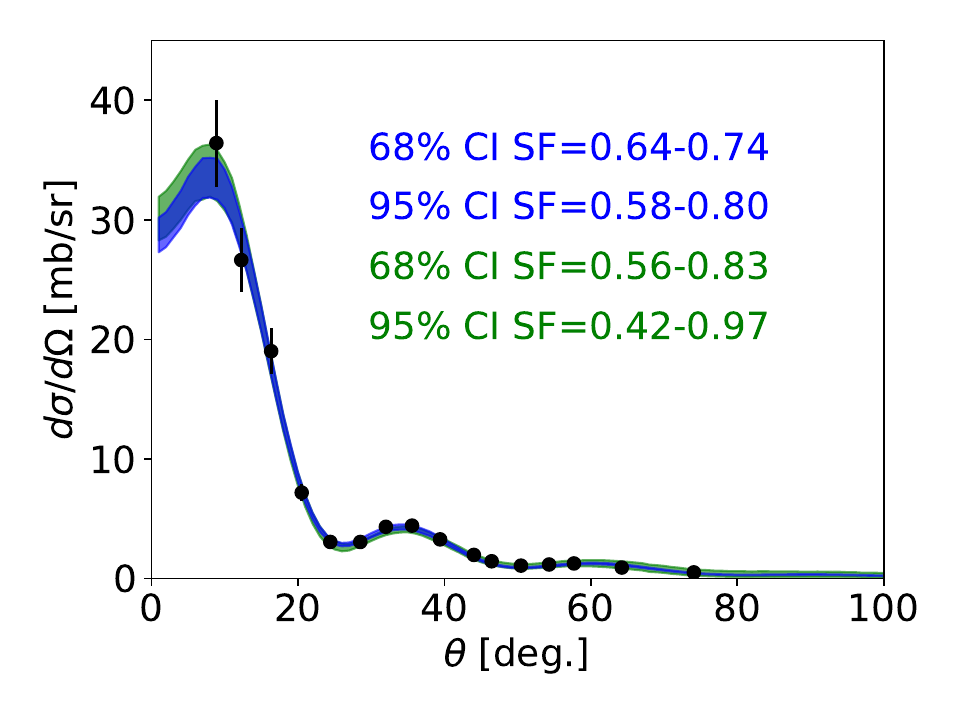}
	\caption{Angular  distribution for $^{48}$Ca$(d,p)^{49}$Ca(g.s.) at 19 MeV scaled to reproduce the first four forward data point, with the corresponding scaling factors (SFs) and their uncertainties. The shaded blue   band corresponds  to the 68\% credible intervals  respectively obtained with optical potentials derived from the same sample of the KDUQ posterior distribution. The green band is obtained using the KDUQ-n posterior distribution for $U_{nA}$ and KDUQ posterior distribution for $U_{pA}$ and $U_{pB}$.  These predictions are compared with data from Ref.~\cite{ca-dp-data}.}
	\label{fig:datadp}
\end{figure}

We now use these parameter posterior distributions and propagate the uncertainties to the $^{48}$Ca$(d,p)^{49}$Ca(g.s.) reaction at beam energy $E_{lab}=19$ MeV.
In Fig.~\ref{fig:datadp}, we show the predicted credible intervals for the corresponding transfer angular distributions: we compare the results obtained with the original KDUQ (blue bands) and those obtained when the neutron interaction is replaced by KDUQ-n (green bands). In these calculations, we fix the neutron bound state using the STD geometry (discussed in Section~\ref{sec:stat}). Fig.~\ref{fig:datadp} already includes the scaling by the SF, taking into account both the optical potential parameter uncertainties propagated in the ADWA model and the experimental error on the transfer data. 
This is done  by adding in quadrature the  errors   ${\varepsilon}_{opt}$ associated with the optical potential and 
$\bar{\varepsilon}_{SF}$ resulting from the fitting procedure to the transfer data. The uncertainty ${\varepsilon}_{opt}$ is  the standard deviation of the  416 SFs minimizing the $\chi^2$ obtained from the transfer data and the  theoretical predictions, e.g., obtained with each KDUQ sample.  
The variance  $\bar{\varepsilon}_{SF}^2$  is calculated by  averaging the variances on the SFs associated with each of the 416 fits.  Further tests have shown that the total uncertainty  on the SF is completely dominated by  ${\varepsilon}_{opt}$ for transfer data errors  $\lsim  10\%$, i.e., the total errors on the SFs are the same regardless of we include $\bar{\varepsilon}_{SF}$  or not.

The theoretical predictions for $\frac{d\sigma}{d\Omega}$ agree well with the data. At the $1\sigma$ level, the relative half-width  $\varepsilon_{68\%}$ Eq.~\eqref{halfwidth_eq}
is $5$\% ($16$\%) when using KDUQ (KDUQ-n).   These uncertainties can be compared with the $20$\% full-width uncertainty shown in green in Fig.~3 of \cite{catacora2023} for the same reaction, at slightly higher beam energy. Note that the work in \cite{catacora2023} uses local potentials with mock data (with $10$ \% error) and systematically renormalizes the likelihood, effectively increasing the error on both proton and neutron optical potentials. {These results demonstrate the benefit of a global parametrization: although there are no  $^{48}$Ca elastic angular distributions in the data corpus used to calibrate the KDUQ parameters, the uncertainties on the transfer predictions are reliable (of the Ca isotopes, the KDUQ data corpus includes only $^{40}$Ca elastic angular distributions).}

\begin{table}
	\centering{}
	\begin{tabular}{cc||c|c}
		uncertainties & final bound state &KDUQ original& KDUQ-n  \\\hline \hline
		\multirow{2}{*}{only scatt. states}& STD  & $5\%$ (SF 6\%)&$13\%$ (SF 16\%)\\
		& KDUQ-real  &$5\%$ (SF 5\%)&$13\%$ (SF 16\%)\\ \hline
		only bound state & KDUQ-real &$4\%$ (SF 4\%)&$22\%$ (SF 23\%)\\\hline
		both scatt. states and bound state & KDUQ-real &$5\%$ (SF 5\%)&$24\%$ (SF 20\%)\\
	\end{tabular}
	\caption{Relative  half-width  $\varepsilon_{68\%}$ Eq.~\eqref{halfwidth_eq} for $^{48}$Ca$(d,p)^{49}$Ca(g.s.) at 19 MeV and, in parentheses, the corresponding relative half-width of the $1\sigma$ credible interval of the extracted SFs using the data from Ref.~\cite{ca-dp-data}. We consider the KDUQ original samples and the rescaled KDUQ posterior KDUQ-n (discussed in the text). Results are obtained when propagating only the uncertainties due to the optical potentials (first two  lines),  only the uncertainties due to the single-particle potential (third line) and both uncertainties (last line). }\label{TableUQ}
\end{table}

To complement our analysis, we also study the uncertainties associated with the single-particle potential used to generate the final  bound state. We quantify its uncertainties using the geometry of the real part of the KDUQ interaction (KDUQ-real), as discussed in Section~\ref{sec:stat}. We consider various cases in Table~\ref{TableUQ} for which we compute the relative  half-widths  of the transfer angular distribution $\varepsilon_{68\%}$ and  of the extracted SF, which also account for the errors on the experimental transfer data.  Specifically, we  investigate if  the geometry of the single-particle potentials, used to generate the final state,  has an impact on the relative uncertainties  due to the optical potentials in transfer observables. First, we include only uncertainties in the optical potential, for two choices of single-particle potentials. In the first two lines of Table~\ref{TableUQ}, we compare the results obtained when using  the  STD single-particle potential ($r_R=1.25$ fm and $a_R=0.65$ fm) and the mean values of the real radius and diffuseness of KDUQ-real ($r_R=1.17$ fm and $a_R=0.689$ fm). 
One can see that the relative uncertainties stay rather constant,  showing that the magnitude of relative uncertainties due to the optical potentials do not depend strongly on the geometry of the single-particle potential in the final state.

Then, we evaluate the uncertainties due to the choice of single-particle potentials, using the geometry of the real volume term of 416 KDUQ and KDUQ-n samples. In the KDUQ case, the uncertainty half-widths~$\varepsilon_{68\%}$ are below 4\% for both the peak of the distribution and the extracted SF, indicating that the KDUQ posterior distributions for $r_R$ and $a_R$ are quite constrained. {The magnitude of the uncertainties on the peak of the transfer cross sections are similar to those on the ANC squared (see Section~\ref{sec:stat}), suggesting that this reaction is  quite peripheral.} As expected, the uncertainties when using the KDUQ-n samples are larger. Interestingly, they are scaled  by roughly the same factor  $\sqrt{38} \sim 6$, as the one used to scale the  KDUQ uncertainties (compare left and right column of the third line). This suggests that the uncertainties in the single-particle radius and diffuseness seem to propagate linearly to transfer observables. Finally, we include both the uncertainties due to the single-particle potential and the optical potentials  (last line). {Contrary to what was found in \cite{catacora2023}, in both cases KDUQ original and KDUQ-n, the total uncertainties cannot be deduced simply by summing in quadrature the two source uncertainties, hinting at the presence of strong correlations.  
	{These  correlations  are due to the interplay between the extension of the single-particle wavefunction, and the range of the real part of the neutron-target optical potential. }
	Although in \cite{catacora2023} the ANC-squared is explicitly used as a constrain, the single-particle potential parameter sampling is independent of the sampling of the optical potential parameters, whereas in this work, KDUQ-real used for the single-particle potential is perfectly correlated to KDUQ (or KDUQ-n) used for the optical potentials.}

Having established a realistic foundation for the uncertainty estimates of the angular distributions of the $^{48}$Ca$(d,p)^{49}$Ca(g.s.) reaction at $E_{lab}=19$ MeV, for which we can compare to experimental data, we now explore how these uncertainties change with beam energy and how they evolve with various properties of the final bound state. {In this exploration,  no uncertainty quantification is included for the bound state interactions - the mean field that binds the neutron in the final state is kept fixed. We vary either the kinematics or  the structure of the final state, and take the optical potential parameters from the same original KDUQ posterior distributions.  This is done in order to show how the same 
	parameter posteriors for optical potentials propagate through the model differently, depending on the details of the reaction. Obviously, because we are not including the additional error in KDUQ-n, nor the uncertainty in the bound state interaction, the overall magnitude of the uncertainty estimates shown in Sections~\ref{sec:eb} and~\ref{sec:spproperties} are underestimated. It is their variation with beam energy or single-particle properties that matters.}

\subsection{Uncertainties in transfer reactions with  beam energy}
\label{sec:eb}

\begin{figure}
	\centering
	\includegraphics[width=0.48\linewidth]{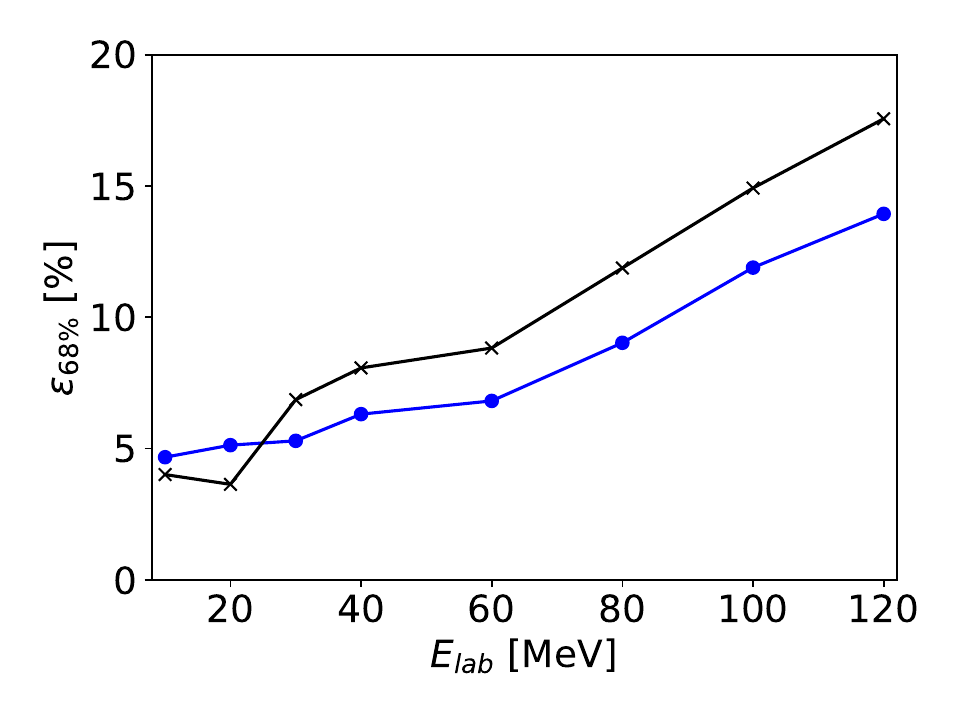}
	\caption{Relative half-width $\varepsilon_{68\%}$ Eq.~\eqref{halfwidth_eq}   for $^{48}$Ca$(d,p)^{49}$Ca(g.s.), as a function of the beam energy. In blue are the results obtained with  nucleon-nucleus interactions needed for the ADWA calculations derived consistently from the same KDUQ sample. The black line corresponds to the situation where all three interactions are derived from different KDUQ samples.  }
	\label{fig:ebeam}
\end{figure}

We first analyze the dependence of the uncertainties  on the beam energy. For this study, we keep the final bound state fixed using the STD single-particle geometry as described in Section~\ref{sec:data},  and take all optical potential posteriors  from the original KDUQ parametrization. 
The relative half-width $\varepsilon_{68\%}$ Eq.~\eqref{halfwidth_eq}, evaluated at the peak of the transfer angular distribution for $^{48}$Ca$(d,p)^{49}$Ca(g.s.), are shown in Fig.~\ref{fig:ebeam}.
There is no convolution with the experimental error on the transfer  data in this plot; only the theoretical uncertainties are considered.

We first consider ADWA calculations using all three optical potentials derived consistently from the same KDUQ sample (blue line). We find that 
the relative  half-width $\varepsilon_{68\%}$ Eq.~\eqref{halfwidth_eq}
increases with the beam energy\footnote{We do not compute transfer cross sections beyond $E_b=120$ MeV as the cross sections then become forbiddingly small to measure. }. This can be explained by the fact that  at higher beam energy, transfer observables become more sensitive to the short-range part of  scattering wave functions, which are typically {less}  constrained by  observables used to calibrate optical potentials.  This explanation was verified by comparing the relative uncertainties obtained when computing the short-range contribution to the radial integral of the T-matrix Eq.~\eqref{T_ad}, i.e., considering only $d$-$^{48}$Ca distances smaller than $R<r_R*48^{1/3}$,  to the uncertainties associated with the long-range contribution to the T-matrix.

To investigate the importance of correlations in the uncertainties of the optical potentials, we consider ADWA calculations using   optical potentials derived from different  KDUQ samples (black line). For almost all beam energies, the relative uncertainties are slightly larger than in the previous results, where all potentials were derived consistently from the same KDUQ sample.  {At the highest beam energies studied, the shift in $\varepsilon_{68\%}$ is about $25$\%.}

\subsection{Dependence on the properties of the final bound state} 
\label{sec:spproperties}

\begin{figure}
	\centering
	\includegraphics[width=0.52\linewidth]{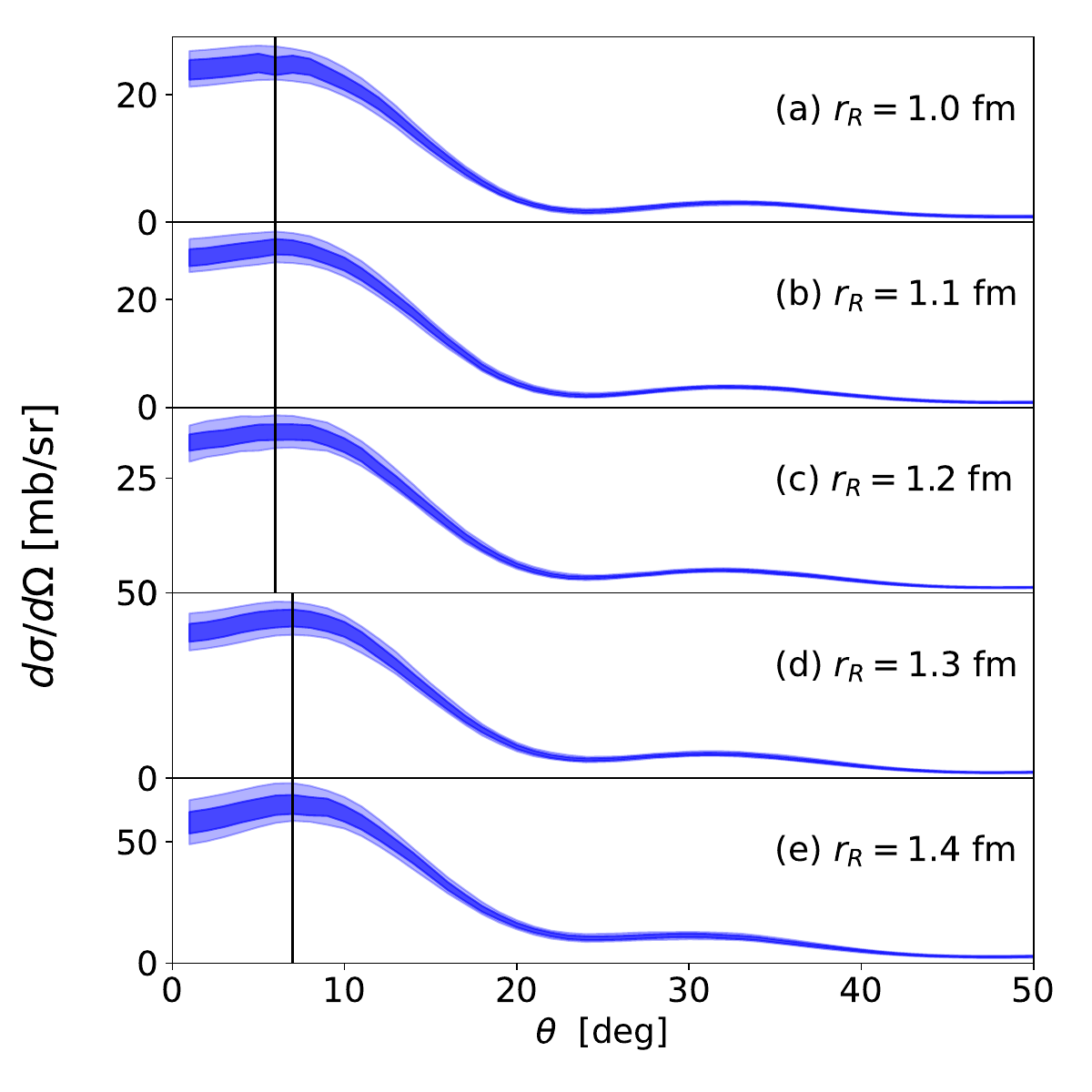}
	\includegraphics[width=0.46\linewidth]{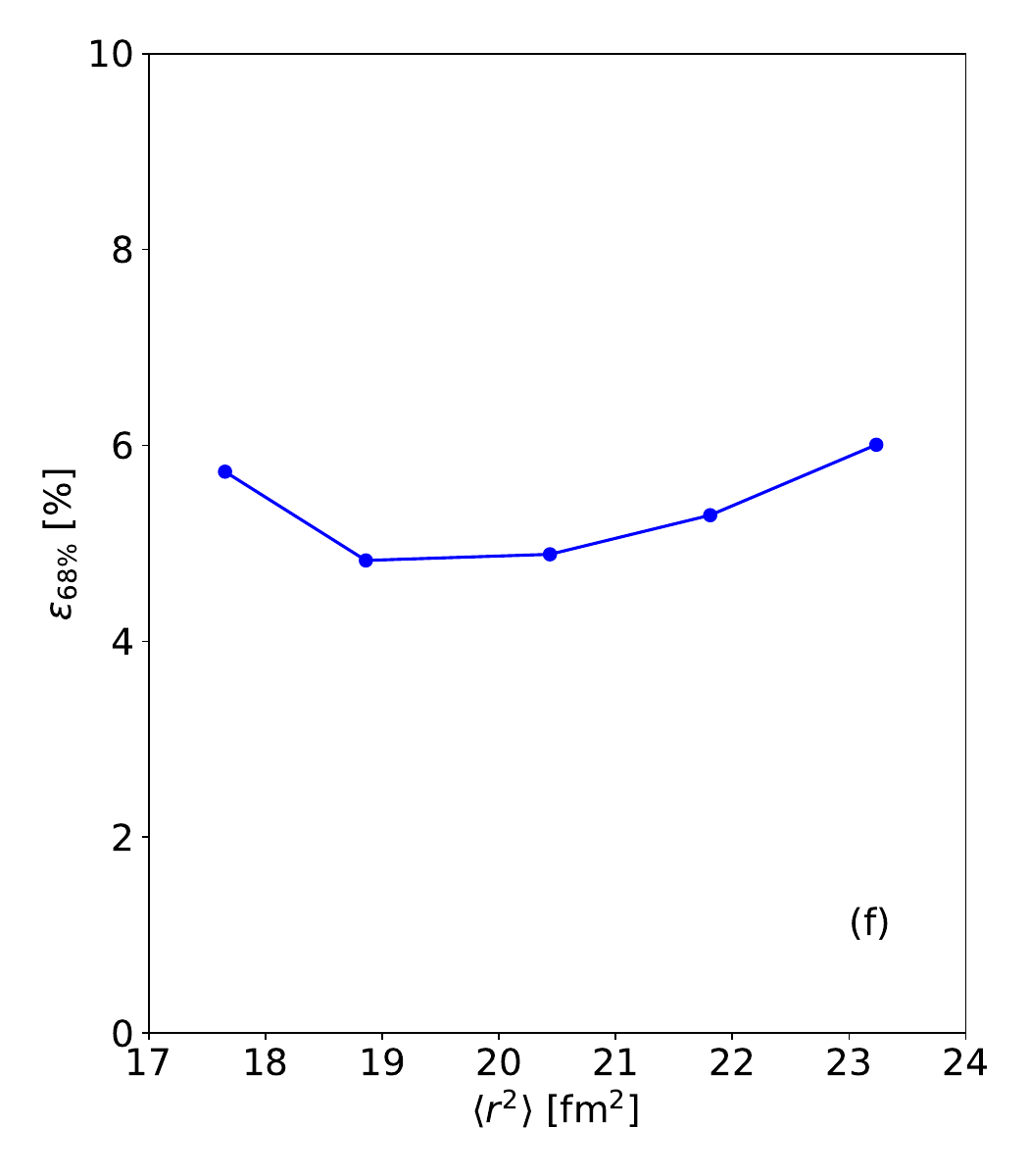}
	\caption{(a-e) Transfer angular distributions for $^{48}$Ca$(d,p)^{49}$Ca(g.s.) at 23 MeV for a range of single-particle radii and (f) the relative  half-width  $\varepsilon_{68\%}$ Eq.~\eqref{halfwidth_eq} as a function of the squared of the single-particle r.m.s radius  $\langle r^2\rangle$. The vertical black lines in panels (a-e) represent the position of the peaks of the transfer distribution $\theta_{max}$.}
	\label{fig:radius}
\end{figure}

Next, we consider the effects of different properties of the final bound state, namely 
the dependence of the uncertainty with the r.m.s. radius squared $\langle r^2\rangle$, the  angular momentum $l$, the number of nodes $n_r$ and the separation energy $S_n$.

We first consider the effect of the single-particle potential  radius $r_R$ used to generate the final bound state wave function on the reaction $^{48}$Ca$(d,p)^{49}$Ca(g.s)  at 23 MeV. We take the original $n$-$^{48}$Ca bound wave function, in a $1p3/2$ orbital, and vary the mean field radius parameter in STD in the range $r_R=1.0-1.4$~fm, along with the depth to reproduce the same separation energy.
The results are shown in Fig.~\ref{fig:radius}: the transfer angular distributions for a range of single-particle potential radii are shown (on the left, panels a-e) and the relative  half-width $\varepsilon_{68\%}$ Eq.~\eqref{halfwidth_eq} as a function of the r.m.s. single-particle radius squared  (on the right, panel f). The same message is relayed when plotting the uncertainty estimates as a function of the ANC squared.

We find that the diffraction pattern of the transfer angular distributions do not change significantly with radius. Expectedly, the magnitude of the transfer cross section increases with the single-particle potential radius $r_R$. Since these reactions are primarily peripheral, they scale with the ANC squared, which in turn is directly related to the r.m.s. radius squared. However, the percent uncertainty remains roughly constant and small, similar to what was observed in Table~\ref{TableUQ}.  Further tests have shown that the same conclusions, i.e., independence of the shape, larger magnitude, small and constant uncertainties for the transfer cross sections, can be drawn when increasing the diffuseness $a_R$.

\begin{figure}
	\centering
	\includegraphics[width=\linewidth]{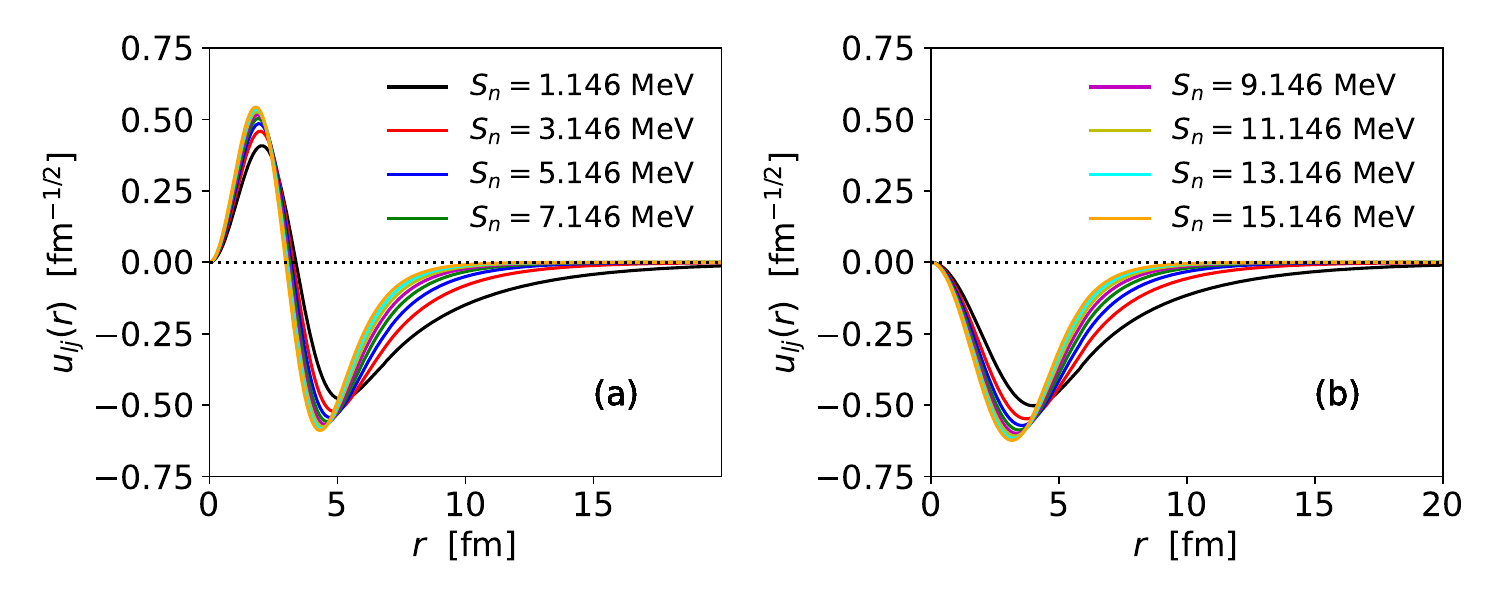}
	\caption{$^{48}$Ca s.p. wave function for a $n$ in a (a) $1p3/2$ and (b) $0p3/2$ states  reproducing various separation  energies $S_n=1.146$-$15.146$ MeV.}
	\label{fig:sn1}
\end{figure}
\begin{figure}
	\centering    
	\includegraphics[width=0.48\linewidth]{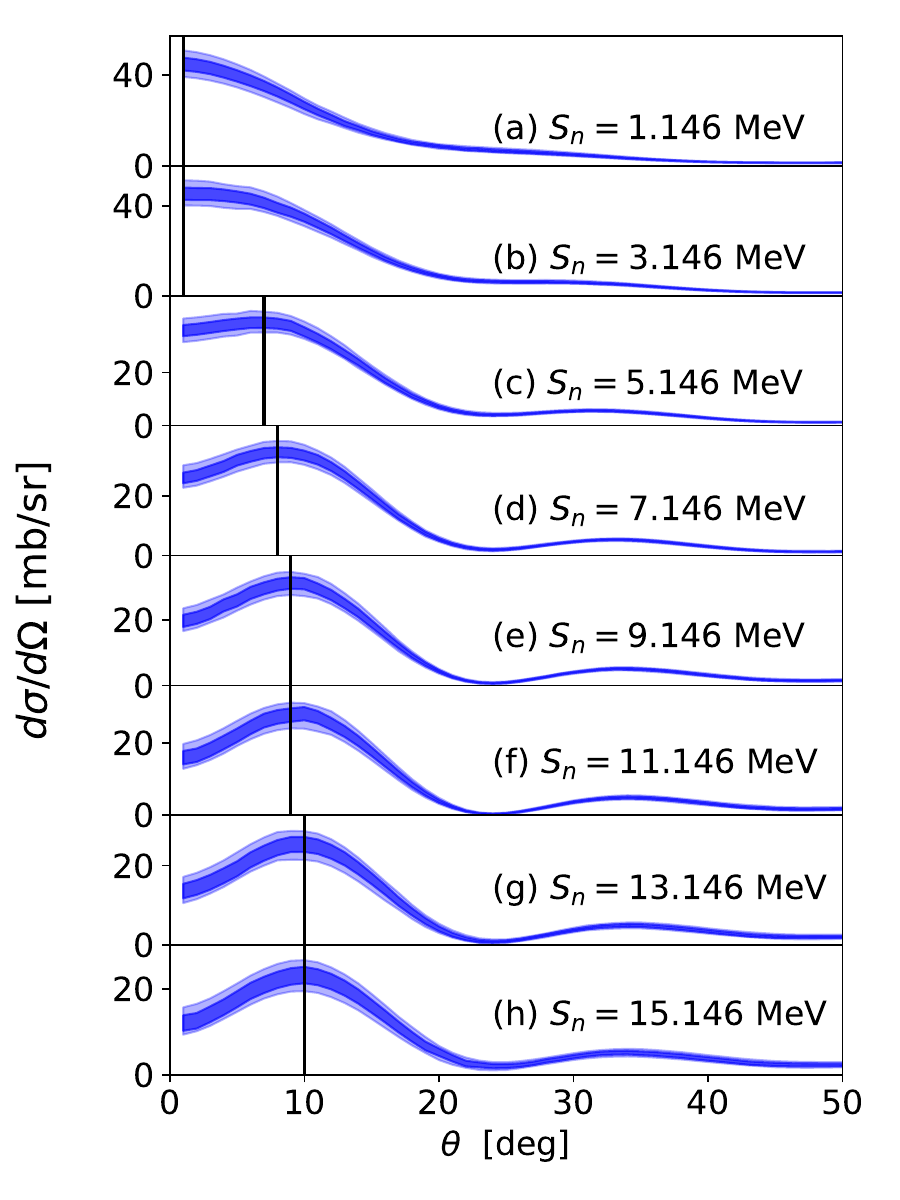}    
	\includegraphics[width=0.48\linewidth]{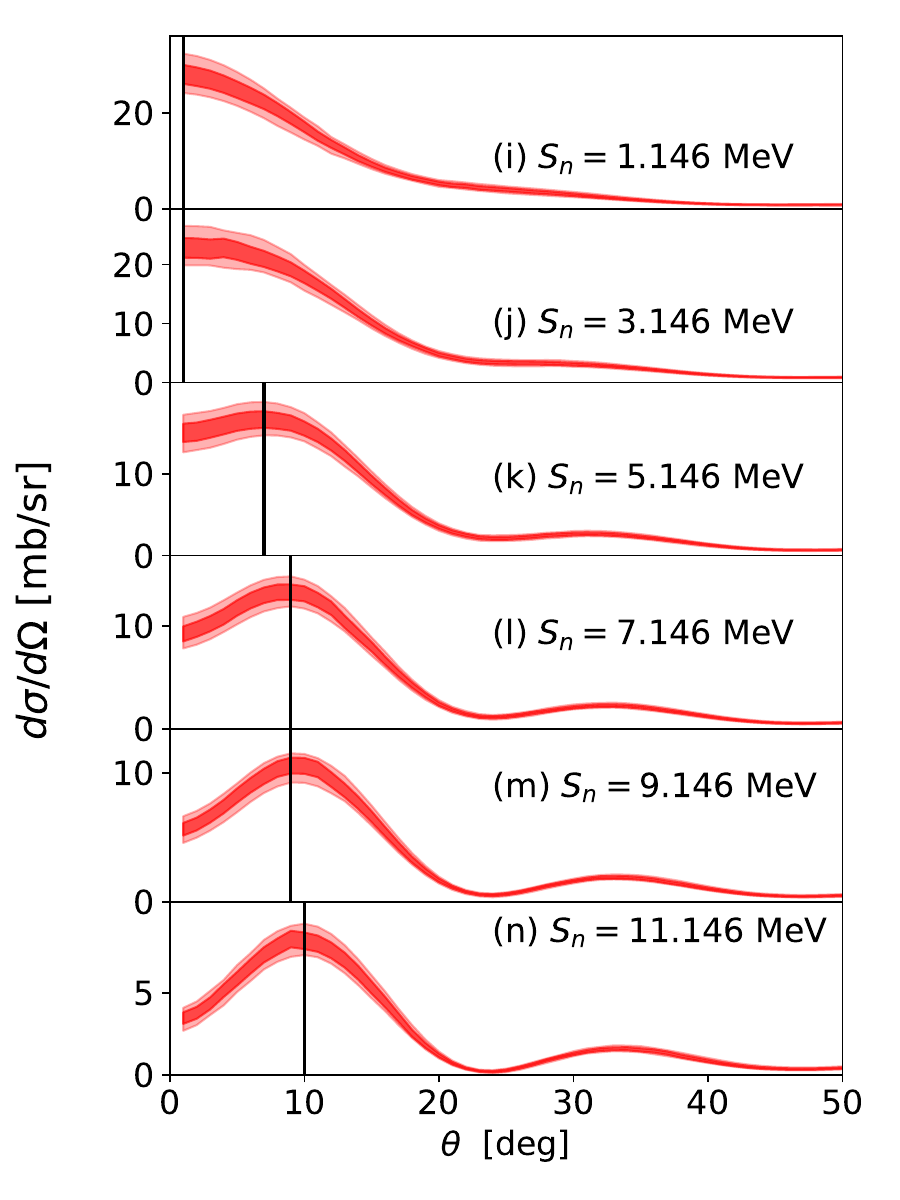} 
	\caption{Transfer angular distributions  for $^{48}$Ca$(d,p)^{49}$Ca(g.s.) at 23 MeV, for different wave function shown in Fig. \ref{fig:sn1}. Panels (a-h) (resp. (i-n))  are obtained with  $^{48}$Ca s.p. wave function for a $n$ in a $1p3/2$ (resp. $0p3/2$).  The vertical black lines in panels (a-h) represent the position of the peaks of the transfer distribution $\theta_{max}$.}
	\label{fig:sn2}
\end{figure}

Next we consider the dependence on the separation energy of the final state, of the uncertainty for the transfer cross section due to  the KDUQ uncertainty. We fix the STD geometry for the neutron single-particle potential to the original values ($=1.25$ fm and $a_R=0.65$ fm) and  adjust the depth of this interaction to reproduce the neutron separation energies $S_n=1.146$-$15.146$~MeV  in the $1p3/2$ wave. We repeat the procedure considering a bound state in a $0p3/2$ orbital. The corresponding wave functions are shown in Fig.~\ref{fig:sn1}(a) and~(b). The lower the separation energy, the more extended is the single-particle wave function. The resulting $(d,p)$ angular distributions  in Fig.~\ref{fig:sn2}, panels (a-h) (resp. (i-n) ) are obtained with a final bound state in the $1p3/2$ (resp. $0p3/2$) wave. In both cases, the peak of the angular distribution shifts to large angles for larger separation energy, as this directly increases the Q-value for the reaction and the momentum matching. The magnitude of the cross section is also affected: the cross sections are larger for bound states with smaller separation energies. This is due to   the spatial extension of the final bound-state wave function.

\begin{figure}
	\centering    
	\includegraphics[width=0.48\linewidth]{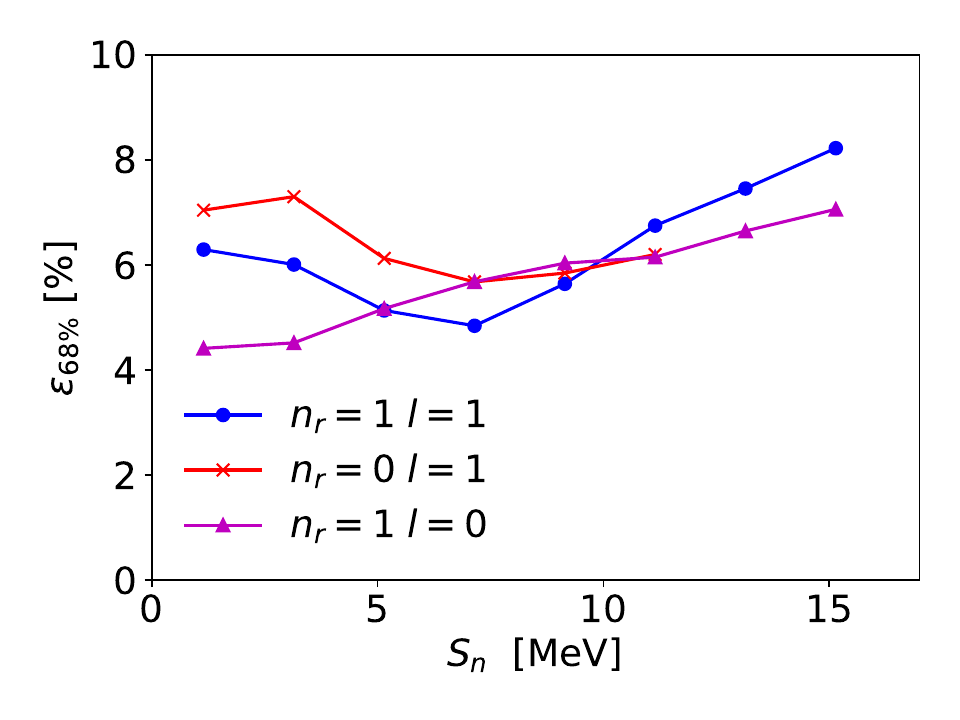}    
	\caption{Relative  half-widths $\varepsilon_{68\%}$ Eq.~\eqref{halfwidth_eq} for  various cases: the blue dots correspond to transfer cross sections populating a $1p3/2$ state, the red crosses to the population of a $0p3/2$ state and the magenta triangles to  the population of a $1s1/2$ state. The corresponding single-particle wave functions and transfer observables are plotted in Figs.~\ref{fig:sn1} and~\ref{fig:sn2}. }
	\label{fig:sn4}
\end{figure}

The resulting relative  half-width $\varepsilon_{68\%}$ Eq.~\eqref{halfwidth_eq} are summarized in Fig.~\ref{fig:sn4}. Here we include not only the uncertainties due to the optical potentials for transfer observables populating a bound state in the $1p3/2$ orbital (blue) and in the $0p3/2$ orbital (red), but also in the $1s1/2$ orbital (magenta). The uncertainties remain small (below $10$\%), regardless of  separation energy, the number of nodes $n_r$ and the angular momentum $l$ of the neutron orbital in the final state.


\section{Conclusions}\label{sec:ccl}

In this work, we perform a systematic study of  parametric uncertainties in $(d,p)$ reactions, and their sensitivity to the kinematics of the reaction, as well as to the properties of the final bound state. The results were obtained using the posterior distribution of the global optical potential KDUQ, enabling us to study the impact of optical potential correlations on transfer observables.

By first analyzing a realistic case, we show that  for  proton scattering off the doubly-magic spherical nucleus $^{48}$Ca, the elastic-scattering cross sections predicted with the KDUQ global optical potential  reproduce well the available scattering data. The empirical coverage lies close to the ideal case, i.e, the diagonal, demonstrating the reliability of the uncertainty estimates of the KDUQ optical potential. However, KDUQ does not reproduce well the neutron scattering data on $^{48}$Ca, suggesting that either the error on the data are seriously under-reported or the uncertainties of KDUQ are unrealistically  small. To account for this, we rescale the KDUQ posterior to obtain an ideal empirical coverage at the $1\sigma$ level. By propagating the posterior distributions in a ADWA model, we find that the relative half-width of the $1\sigma$ credible interval at the peak  of the transfer angular distributions is about $5\%$ when using the KDUQ parameters, and 25$\%$ when using the rescaled  KDUQ-n parameters. Moreover, we note that the uncertainties due to the single-particle binding potentials are below 5\% when using KDUQ and 25\% for the rescaled KDUQ-n. 
Interestingly, our results also show that, in transfer observables, the two uncertainties (from the single-particle binding potential and from the optical potentials) do not add in quadrature. {This suggests the impact of correlations between the single-particle binding potential and the optical potential parameters is significant.}

Then fixing the geometry of the single-particle potentials and considering the same KDUQ posterior distribution,  we investigate how  the uncertainties in transfer observables are influenced by the beam energy of the reaction and the properties of the final bound state. Since the same KDUQ posterior parameters are taken in all cases, the different uncertainties do not come from the evolution of KDUQ uncertainties across nuclei or with the beam energy, but are a direct consequence on how uncertainties 
propagate through the model differently, depending on the details of the reaction. 

We show  that at higher beam energy, the uncertainties in transfer observables increase. This is a direct consequence of the radial range probed by transfer reactions at various beam energies: transfer reactions at higher beam energies are  more sensitive to the short-range part of the T-matrix, which is not well constrained for optical potentials fitted on elastic observables. 
{We find that the correlations in the optical potentials used to generate the scattering states can change the uncertainty estimate by $20-25$\%.}

We also  investigate how uncertainties due to optical potentials depend on the properties of the final bound state: we vary the geometry of the single-particle potential, the binding energy, the orbital angular momentum and the number of nodes. As expected, the magnitude and the shape of the transfer cross sections change, as they are influenced by the spatial extension of the bound-state wave function, its orbital angular momentum and  the $Q$-value of the reaction. Nevertheless, the relative half-widths $\varepsilon_{68\%}$ Eq.~\eqref{halfwidth_eq} remain {below $10$\%} for all the cases covered in this study. 
{When using KDUQ,} the magnitude of the optical model uncertainties on transfer observables is not strongly dependent on the properties of the bound state.

Although our results are quite general, there were important simplifying assumptions that should be kept in mind.  First, our analysis does not account for  uncertainties associated with the ADWA approximation to the three-body dynamics. In particular, the neglect of the remnant term is likely to become inaccurate for reactions on light nuclei or populating halo final states.
Second, we use the same KDUQ posterior distributions for  all cases. 
This assumption likely leads to an underestimation of parametric uncertainties  for transfer reactions at higher beam energies, since  KDUQ's relative uncertainties are larger at higher  energy (as illustrated in the volume integral in Fig. 13 of Ref.~\cite{KDUQ}). 
Finally, the KDUQ posterior distribution we chose is likely unrealistic for nuclei with low separation energies, since these isotopes are far from the valley of stability. No data on unstable nuclei {or deformed nuclei} were included in the calibration of KDUQ and, therefore, one expects the uncertainties to grow substantially as we move to more exotic territory.

This systematic study enables us to draw three important take-away points.  First, for well constrained potentials, such as KDUQ, small uncertainties in transfer observables can be expected, typically around 5\%-10\%. Second, there are significant correlations between the single-particle potential and optical potential parameters that impact the estimated uncertainties on the transfer. This argues for a framework where  bound state data and  scattering data can both be used to constrain the same interaction consistently, something that is obtained by imposing the dispersive relation~\cite{DICKHOFF2019252}. We should expect that, because in the dispersive optical model bound state data is used to the potential,  it may lead to a better constrained off-shell part of the T-matrix, hence reducing the uncertainties on reaction observables that do not solely depend on the on-shell properties (such as elastic scattering). Third, our results show that the relative uncertainty estimates of transfer angular distributions are not sensitive to detailed properties of the neutron orbital in the final state populated by the transfer reaction.

\clearpage
\section*{Conflict of Interest Statement}

The authors declare that the research was conducted in the absence of any commercial or financial relationships that could be construed as a potential conflict of interest.

\section*{Author Contributions}
All authors have made a substantial, direct, and intellectual
contribution to the work and approved it for publication.

\section*{Funding}
The work of F. M. N. was in part supported by the U.S. Department of Energy grant DE-SC0021422 and National Science Foundation CSSI program under award No. OAC-2004601 (BAND Collaboration).
\section*{Acknowledgments}
C. H. and F. M. N. thank  Kyle Beyer, Manuel Catacora-Rios,  Garrett King {and other members of the few-body reaction group at MSU} for their careful reading of the manuscripts and their comments.
C. H. thanks Cole Pruitt  and Gregory Potel for  interesting discussions regarding the results of this work.   We gratefully acknowledge computational support from iCER at Michigan State University.

\section*{Data Availability Statement}
The raw data supporting the conclusion of this article will be
made available by the authors, without undue reservation.
\bibliographystyle{Frontiers-Vancouver} 
\bibliography{reactions}

\end{document}